\documentclass[journal=jacsat,manuscript=article]{achemso}
\usepackage[version=3]{mhchem} 
\usepackage{pdfpages}

\author{M.~C. Giordano}
\affiliation{\'Ecole Polytechnique F\'ed\'erale de Lausanne, School of Engineering,  Institute of Materials, Laboratory of Nanoscale Magnetic Materials and Magnonics, 1015 Lausanne, Switzerland}
\author{K. Baumgaertl}
\affiliation{\'Ecole Polytechnique F\'ed\'erale de Lausanne, School of Engineering,  Institute of Materials, Laboratory of Nanoscale Magnetic Materials and Magnonics, 1015 Lausanne, Switzerland}
\author{S.~R. Escobar Steinvall}
\affiliation{{\'Ecole Polytechnique F\'ed\'erale de Lausanne, School of Engineering,  Institute of Materials, Laboratory of Semiconductor Materials, 1015 Lausanne, Switzerland}}
\author{J. Gay}
\affiliation{\'Ecole Polytechnique F\'ed\'erale de Lausanne, School of Engineering,  Institute of Materials, Laboratory of Nanoscale Magnetic Materials and Magnonics, 1015 Lausanne, Switzerland}
\author{M. Vuichard}
\affiliation{\'Ecole Polytechnique F\'ed\'erale de Lausanne, School of Engineering,  Institute of Materials, Laboratory of Nanoscale Magnetic Materials and Magnonics, 1015 Lausanne, Switzerland}
\author{A. Fontcuberta i Morral }
\affiliation{{\'Ecole Polytechnique F\'ed\'erale de Lausanne, School of Engineering,  Institute of Materials, Laboratory of Semiconductor Materials, 1015 Lausanne, Switzerland}}
\altaffiliation{\'Ecole Polytechnique F\'ed\'erale de Lausanne, School of Basic Sciences,  Institute of Physics, 1015 Lausanne, Switzerland}
\author{D.~Grundler}
\affiliation{\'Ecole Polytechnique F\'ed\'erale de Lausanne, School of Engineering,  Institute of Materials, Laboratory of Nanoscale Magnetic Materials and Magnonics,  1015 Lausanne, Switzerland}
\altaffiliation{\'Ecole Polytechnique F\'ed\'erale de Lausanne, School of Engineering,  Institute of Microengineering, 1015 Lausanne, Switzerland}
\email{dirk.grundler@epfl.ch}
\phone{+41 21 693 38 52}
\fax{}

\title {Plasma-enhanced atomic layer deposition of nickel nanotubes with low resistivity and coherent magnetization dynamics for 3D spintronics}

\abbreviations{}
\keywords{American Chemical Society, \LaTeX}

\begin{document}
\begin{abstract}
We report plasma-enhanced atomic layer deposition (ALD) to prepare conformal nickel thin films and nanotubes by using nickelocene as a precursor, water as the oxidant agent and an in-cycle plasma enhanced reduction step with hydrogen. The optimized ALD pulse sequence, combined with a post-processing annealing treatment, allowed us to prepare 30 nm thick metallic Ni layers with a resistivity of 8 $\mu\Omega$cm at room temperature and good conformality both on the planar substrates and nanotemplates. Thereby we fabricated several micrometer-long nickel nanotubes with diameters ranging from 120 to 330 nm.
We report on the correlation between ALD growth and functional properties of individual Ni nanotubes characterized in terms of magneto-transport and the confinement of spin wave modes.
The findings offer novel perspectives for Ni-based spintronics and magnonic devices operated in the GHz frequency regime with a 3D device architecture.
\end{abstract}

\section{Introduction}

Magnetic nanostructures find applications in data storage devices \cite{Makarov2016}, magnetic sensors \cite{Koh2009,Mccray2009} and biomedical applications \cite{Chen2015}. The demand of high-density technologies \cite{Parkin2008} and the need of new applications are driving the expansion of nanomagnetism towards three-dimensional (3D) device architectures\cite{Streubel2016,Fernandez-Pacheco2017}. Here, novel physical phenomena arising from complex spin textures become possible. However, the implementation of high-quality 3D nanomagnetic structures on large scales is lacking. A promising path is to magnetically coat 3D nanotemplates, e.g. polymeric scaffolds\cite{Donnelly2015}, self-assembled\cite{Streubel2012} or bottom-up grown\cite{Ruffer2012,Ruffer2014} nanostructures. Therefore, a technique for conformal coatings of high-quality ferromagnetic metals is of uttermost importance to advance 3D nanomagnetism.
Atomic layer deposition (ALD) \cite{Johnson2014,George2010} indeed offers a great potential. Here the material is deposited through self-limiting reactions between vapor phase metal-organic precursors (and co-reactants) and the exposed substrate surface, enabling conformal coatings and thickness control on the atomic scale. For the deposition of numerous oxides, ALD is a well-established technique. For metallic thin films, including ferromagnetic metals, ALD still faces challenges due to the limited number of suitable precursors, the difficulty in reducing the metal cations, and the tendency of metals to agglomerate to islands \cite{Hagen2019,Lim2003}.
Previous attempts to deposit the ferromagnetic metal nickel with ALD included one-step and two-step processes. In the first scenario, the nickel films were deposited directly through the reaction of the nickel precursor and a co-reactant. Amongst them there were molecular hydrogen \cite{Lim2003}, molecular ammonia\cite{Kim2011}, plasma-activated hydrogen\cite{Lee2010} and ammonia \cite{Lee2010,Wang2016}, as well as tert-butilamine\cite{Kerrigan2018}. The use of plasma offered the possibility to lower the deposition temperature and to increase the purity of the deposited material \cite{Profijt2011}. However, the recombination of plasma species limited the conformality of plasma-enhanced ALD (PE-ALD) on high aspect ratio (AR) nanotemplates \cite{Cremers2019,Profijt2011}. For the two-step process, the first step was based on nickel oxide deposition, where the oxidant co-reactant was ozone\cite{Ruffer2012,Daub2007,Wang2012} or water\cite{Chae2002}, and the second step relied on a reduction process, which transformed the oxide into pure nickel. The reduction was performed either within the ALD cycle with hydrogen plasma \cite{Chae2002}, or as post-deposition annealing in molecular hydrogen \cite{Ruffer2012,Ruffer2014}, or by exploiting both processes.
The two-step process ensured, in principle, good conformality on high aspect ratio nanotemplates due to the greater homogeneity of nickel oxide deposition via the thermal ALD growth. The usage of water as co-reactant resulted in Ni coatings with good conformality on nanotemplate substrates with an AR of about 3:1 \cite{Chae2002}. Metallic Ni coatings achieved with ozone as the co-reactant exhibited appreciable surface roughness and inhomogeneities as Ostwald ripening was potentially induced by the annealing temperatures above 400$~^\circ$C \cite{Ruffer2012,Espejo2016}.\\
Nanotubes (NTs) prepared from ferromagnets represent prototypical 3D nanomagnetic structures \cite{Fernandez-Pacheco2017}. They raised significant attention as they support stable flux-closure magnetic states \cite{Stano2018} and avoid the Bloch point structure along the central axis. This results in a fast and controllable reversal process when compared to magnetic stripes \cite{Landeros2007}. In addition, previously unforeseen dynamic effects are possible. Domain walls moving in nanotubes are predicted to avoid Walker breakdown and give rise to Cherenkov-like spin wave emission \cite{Yan2011}. Their curvature generates non-reciprocal spin-wave dispersion \cite{Otalora2016}. Furthermore, the three geometric parameters, namely the length, the inner radius and external radius, offer the possibility to tailor spin-wave confinement. These characteristics make ferromagnetic NTs promising functional objects for 3D integrated spintronics and magnonics \cite{Neusser2009}.
Experimental studies on individual NTs focused so far on the determination of their static magnetic properties, magnetization states and reversal magnetization mechanisms, by techniques such as SQUID magnetometry \cite{Buchter2013}\cite{Vasyukov2018}, cantilever magnetometry \cite{Weber2012}, anisotropic magnetoresistance measurements \cite{Ruffer2012,Ruffer2014,Baumgaertl2016} and x-ray based magnetic imaging \cite{Zimmermann2018,Vasyukov2018}.
Still, the dynamic magnetization of individual ferromagnetic NTs with the radius on the order of 100 nm has not yet been the subject of comprehensive experimental investigations. So far studies were limited to spin-wave resonances of large ensembles of ferromagnetic tubular structures \cite{Wang2005} and of rolled up ferromagnetic layers on semiconductor membranes with micrometric radii\cite{Balhorn2012,Balhorn2013}, where azimuthal interference of long-wavelength magnetostatic spin waves was found. Lenz et al.\cite{Lenz2019} recently reported the magnetization dynamics of an individual Fe nanorod embedded in a carbon nanotube, studied by ferromagnetic resonance (FMR) and Brillouin light scattering (BLS). Dynamics of individual nanotubes \cite{Yan2011,Otalora2016} remain to be explored experimentally.\\
Atomic layer deposition (ALD) has selectively been used for the fabrication of ferromagnetic Ni NTs by coating GaAs nanowires \cite{Ruffer2012,Ruffer2014}. Weber {\em et al.} \cite{Weber2012} demonstrated three stable magnetization states in individual ALD-grown Ni nanotubes measured by cantilever magnetometry. Rueffer {\em et al.} \cite{Ruffer2014} measured an anisotropic magnetoresistance (AMR) effect of up to 1.4 $\%$ on such Ni nanotubes when applying fields of 1~T at room temperature. The presented NTs exhibited a rough surface after annealing and their magnetization dynamics were not reported. Wang {\em et al.} \cite{Wang2016} reported PE-ALD to coat high aspect ratio (AR$\sim13:1$) Si nanopillars with Ni. They achieved a low resistivity of 11.8 $\mu\Omega$cm in planar Ni layers after post-annealing enhancement, but did not report on functional properties of the high aspect ratio coatings.\\
In this work we present the fabrication of Ni NTs by PE-ALD on single-crystalline GaAs nanowires as 3D nanomagnetic structures. Adapting the two-step ALD process of Ref.~\cite{Chae2002} we employed nickelocene as the metallorganic precursor and water as the oxidant agent to first obtain nickel oxide. Subsequently, an in-cycle plasma enhanced reduction process with hydrogen was performed to get metallic nickel. We optimized the ALD sequence to enhance the functional properties and conformality of Ni in the nanotubular shape. We identified the deposition conditions that led to identical coverage on planar substrates and on the nanotemplates with an AR above 15:1. The optimized Ni NTs exhibited, after post-annealing enhancement, smooth surfaces and showed both a low resistivity of roughly 8 $\mu\Omega$cm and a prominent AMR effect. We characterized the magnetization dynamics of individual NTs under microwave irradiation in the GHz frequency regime. We detected a series of resonances by micro-focused BLS which we attributed to spin-wave modes with discrete wave vectors $k$ due to azimuthal confinement. To the best of our knowledge, this is the first experimental investigation of resonant modes in nanoscale ferromagnetic NTs.
Our results prove the high quality of the achieved 3D ferromagnetic coatings, promising to advance the realization and exploration of 3D spintronic device architectures such as the proposed racetrack memory.\cite{Parkin2008}

\section{Results and discussion}

\subsection{Growth of nickel thin films and nanotubes}

We start by reporting on the PE-ALD growth of Ni thin films on planar Si (100) substrates and high aspect ratio nanotemplates. The planar wafers and GaAs nanowires with hexagonal cross-sections (Figure S1), respectively, were coated with a few nanometers of alumina with the aim of growing Ni on the same surface, while focusing on the effect of different substrate geometries and aspect ratios. In order to optimize the composition and morphology we varied the chamber temperature $T$ and the number of metal oxide deposition steps \textit{c} prior to hydrogen-plasma reduction.
Figure 1a shows the dependence of the growth rate of nickel on planar substrates on $T$. Here the number of nickel oxide was set to $c=1$ and the ALD cycle was repeated $n = 1000$ times (inset of Figure 1b). The chamber temperature was varied from 155 $^\circ$C to 175 $^\circ$C. From $T = 155~^\circ$C to $T=165~^\circ$C we observe that the growth rate increases from a value of 0.12 \AA/(\ce{NiCp2} pulse) to a value of 0.33 \AA/(\ce{NiCp2} pulse). It stays constant at 0.33 $\pm$ 0.01 \AA/(\ce{NiCp2} pulse) for temperatures ranging from 165 to 172 $^\circ$C. We attribute this temperature regime to the ALD window. The growth rate is at 0.47 \AA/(\ce{NiCp2} pulse) for $T=175~^\circ$C.
\begin{figure}[h!]
	\includegraphics[width=0.5\textwidth]{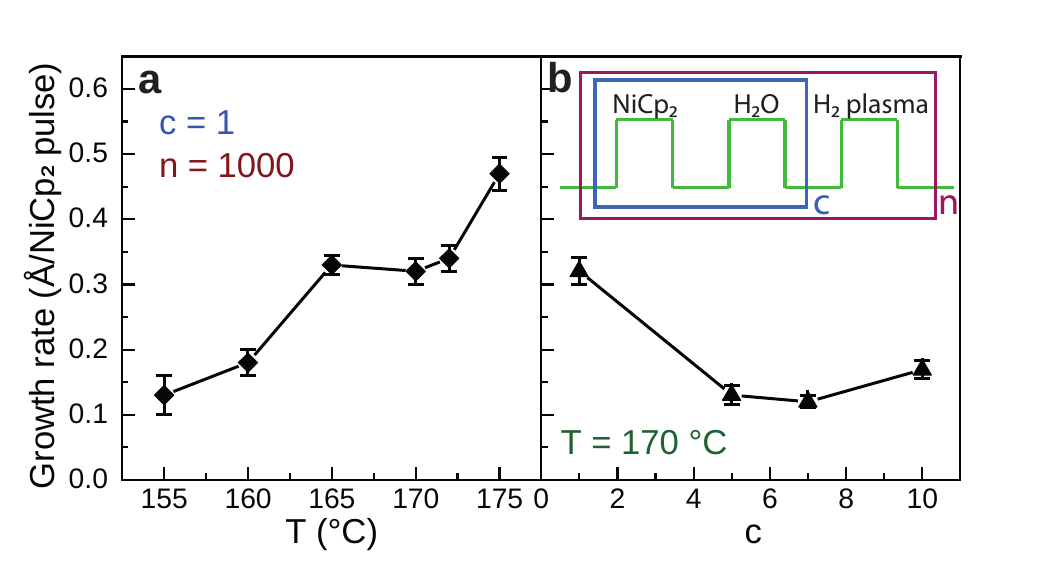}
	\caption{(a) Dependence of the growth rate of Ni on the chamber temperature. Each sample was prepared with $n$ = 1000 cycles. (b) Dependence of growth rate on the number $c$ of \ce{NiCp2}/\ce{H2O} steps in the ALD cycle (inset) for samples deposited at $T = 170~^\circ$C.}\label{Figure 1}
	
\end{figure}

\begin{figure}[h!]
\includegraphics[width=0.7\textwidth]{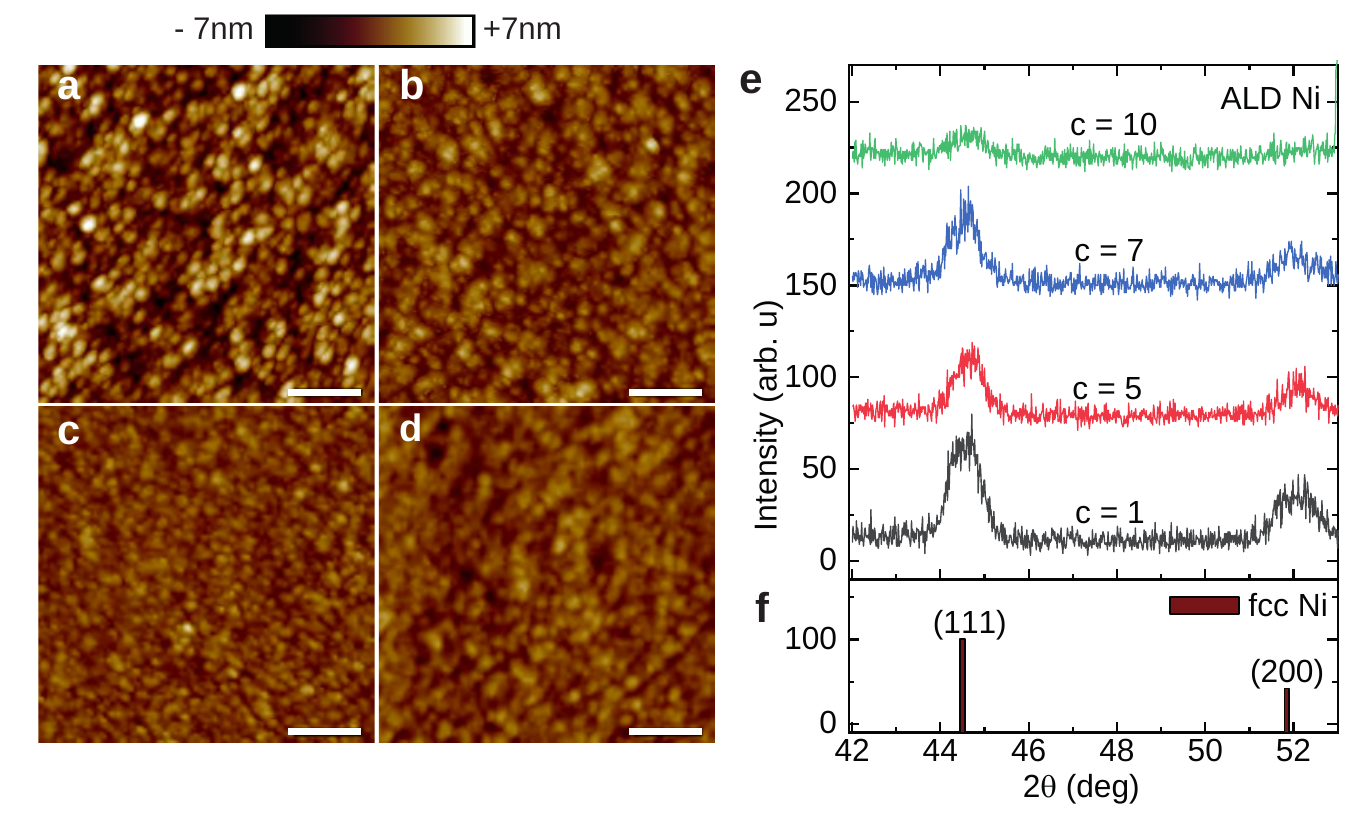}
\caption{Atomic force microscopy performed on a $1~\mu$m$\times 1~\mu$m area of samples with (a) $c=1$, (b) $c=5$, (c) $c=7$, and (d) $c=10$ (scale bars: 200 nm). (e) X-ray diffractograms of $30~\pm2$ nm thick nickel films prepared with $c$ ranging from 1 to 10 (see labels). (f) The bars give the positions of the (111) and (200) reflections of cubic nickel (PDF 00-004-0850) . The thin films were deposited at 170 $^\circ$C on a silicon substrate coated with 5 nm of \ce{Al2O3} and annealed at 350$~^\circ$C.}\label{Figure 2}
\end{figure}

To explore how the growth rate varies at $T = 170~^\circ$C with changing the number of nickel oxide steps $\textit{c}$ in the ALD cycle, we varied the parameter $c$ from 1 to 10, while the ALD cycle $n$ was changed consistently in order to keep the product $c\times n = 1000$. In Figure~\ref{Figure 1}b we report the growth rate estimated on the the number of total $\ce{NiCp2}$ pulses ($c\times n$) as function of the parameter $\textit{c}$, number of steps leading to nickel oxide formation. We observe that the growth rate drops from 0.33~\AA/(\ce{NiCp2} pulse) at $c = 1$ to $0.13$~\AA/(\ce{NiCp2} pulse) at $c = 5$ and $7$. It increases again to 0.17~\AA/(\ce{NiCp2} pulse) for $c=10$. Thin films with a thickness of about 30 nm were prepared with $c$ set to $1,5,7$ and 10, taking into account the values of growth rate reported in Figure~\ref{Figure 1}b. They were then annealed under hydrogen flow (details in Methods).
In the following we discuss the properties of annealed Ni films. Atomic force microscopy (Figure 2a-2d) and X-ray diffraction (Figure 2e) were used to assess the roughness and the crystallographic structure, respectively, of the four planar Ni films. We extracted rms roughness values of 2.1 nm for $c=1$ (Figure~\ref{Figure 2}a), $1.3$~nm for $c$ = 5 (Figure~\ref{Figure 2}b), 0.9 nm for $c=7$ (Figure~\ref{Figure 2}c) and 1.2 for $c = 10$ (Figure~\ref{Figure 2}d). Accordingly, the sample prepared with $c=7$ showed the smoothest surface (compare Figure~\ref{Figure 5}a).

\begin{figure}[h!]
	\includegraphics[width=0.45\textwidth]{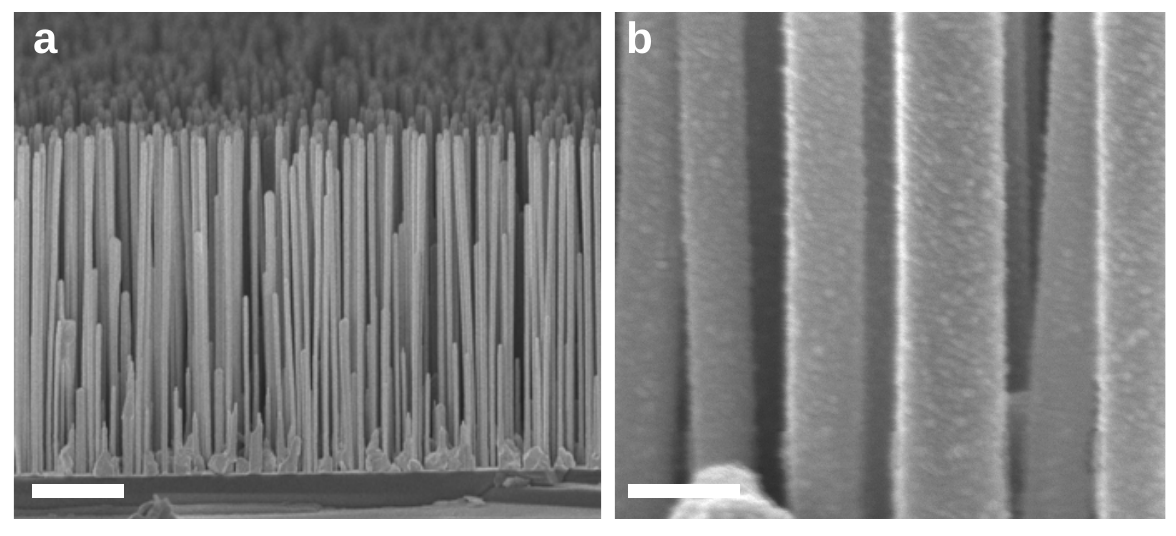}
	\caption{Scanning electron microscopy (SEM) micrographs of (a) a large ensemble of vertical nanowires with ALD-grown Ni shells (scale bar: 1 $\mu$m) and (b) selected central regions of Ni NTs (scale bar: 200 nm).}\label{Figure 3}
\end{figure}

In Figure~\ref{Figure 2}e we report the X-ray diffractograms acquired in a glancing incident angle configuration for the four annealed thin films. The measured peaks are consistent with the (100) and (200) reflections of cubic nickel (PDF 00-004-0850) shown in Figure~\ref{Figure 2}f. The broadening and positions of peaks of the samples prepared with $c=1, 5$ and $7$ suggest polycrystalline Ni films, with no preferential grain orientation. \\
The four depositions were performed simultaneously on Si wafers and GaAs nanowires (both coated with alumina). We employed different set of nanowires, with lengths between 7 and 15 $ \mu$m and diameters ranging from 100 to 270 nm. The aspect ratios of different nanotemplates were calculated following the convention for squared pillars \cite{Cremers2019} and range from 15:1 to 31:1 (Table~S1).
The nanotemplates allowed us to obtain arrays of millions of vertically oriented nickel nanotubes (Figure~\ref{Figure 3}).
\begin{figure}[h!]
	\includegraphics[width=1\textwidth]{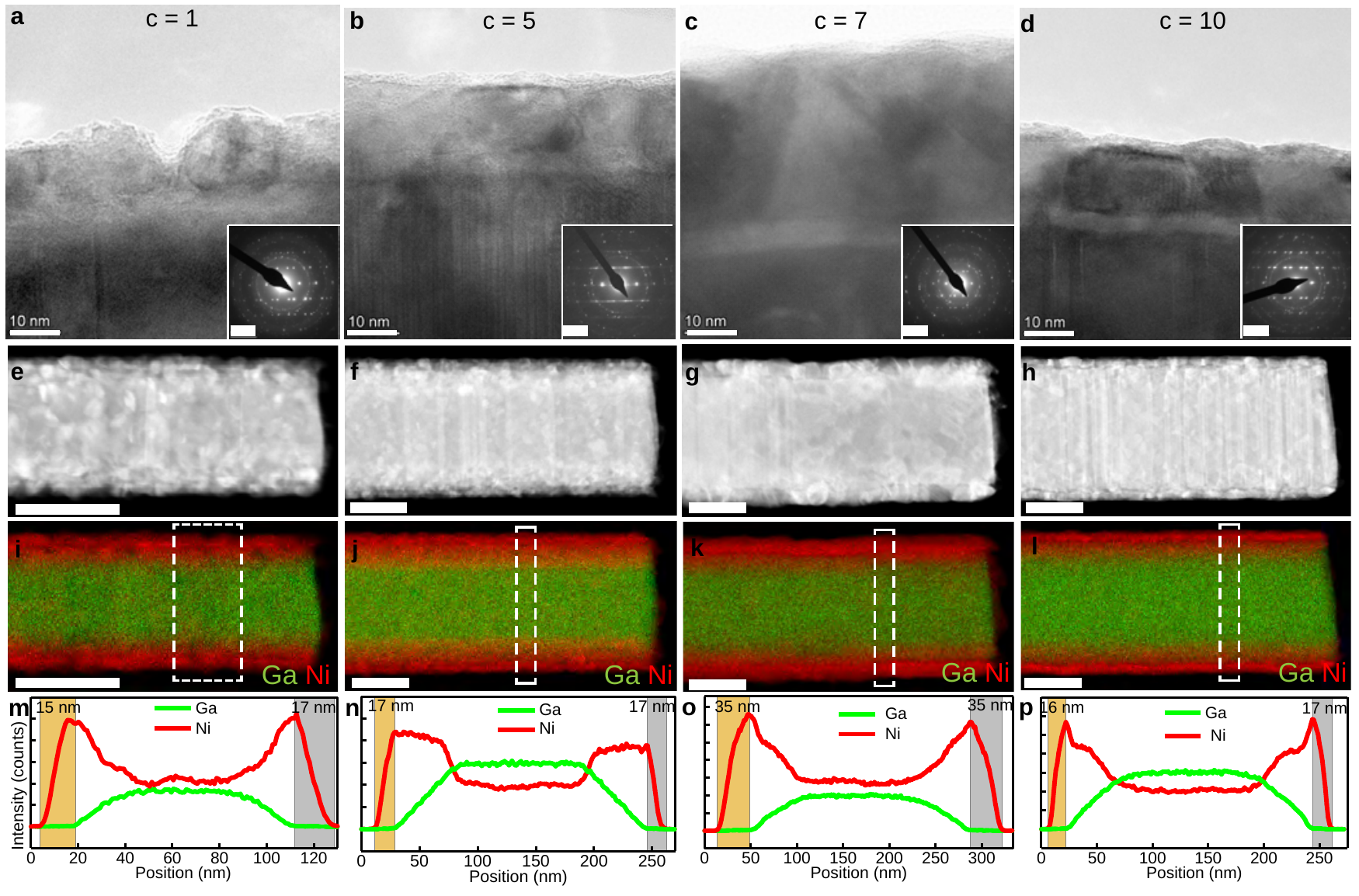}
	\caption{(a-d) HRTEM images (scale bar: 10 nm) of interfaces between GaAs nanowires and the $\ce{Al2O3}$/Ni shells. The insets show SAED patterns (scale bars: 5 nm$^{-1}$). (e-h) HADDAF images (scale bars: 100 nm). (i-l) Distribution of Ga and Ni. (m-p) Elemental analysis of Ni nanotubes on GaAs nanowires. Each column of panels represents one sample. The four samples were prepared with $c$ set to $1, 5, 7$ and $10$ as indicated on top of each column. The chamber (annealing) temperatures were 170 $^{\circ}$C (350$^\circ$C).}\label{Figure 4}
\end{figure}

To investigate the chemical composition and the structure of the Ni NTs we performed transmission electron microscopy (TEM).
Figure~\ref{Figure 4}a to \ref{Figure 4}d shows high-resolution (HR) TEM micrographs with insets of the corresponding selective area electron diffraction (SAED) patterns of the Ni nanotubes prepared with $c= 1,5,7$ and $10$, respectively. The images depict the interface between GaAs nanowires and the ALD-grown \ce{Al2O3}/Ni shells. The Ni coating with $c = 1$ (Figure~\ref{Figure 4}a) exhibits a relatively large peak-to-peak roughness value of about 7 nm. The Ni shells deposited with $c = 5, 7, 10$ (Figure~\ref{Figure 4}b to d) show a continuous coverage and form ferromagnetic nanotubes with small roughness. The roughness variation of the NTs is qualitatively consistent with the characteristics found for the planar Ni films (Figure~\ref{Figure 5}a). No preferential growth direction is observed in any of the samples, all made up of grains of different sizes and orientations. An epitaxial relationship between core and shells was not found. All shells exhibited a surface layer of nickel oxide of a few nanometers, attributed to oxidization in air after the annealing treatment. SAED images contain the typical ring patterns of a polycrystalline texture, which we attribute to the nickel shell, and a single crystal dot pattern of a zinc-blende structure, that we assign to the GaAs core, which was grown along the [111]B directions as previously reported \cite{Uccelli2011}. Starting from the center, the diffraction rings were assigned to the (111), (200), (220) and (311) planes respectively, belonging to the face-centered cubic lattice of Ni crystals.
Using scanning TEM (STEM) we performed high-angle annular dark-field (HAADF) and energy dispersive X-ray spectroscopy (EDS) analysis to study the composition and elemental distribution of the nanostructures. HADDAF images in Figure \ref{Figure 4}e to \ref{Figure 4}h display the edges of NTs which were cleaved and close to the Si(111) substrate. Corresponding element maps of Ga and Ni, as representative elements of inner core and outer shell, are shown in Figure \ref{Figure 4}i to \ref{Figure 4}l (and Figure~S1). Figure \ref{Figure 4}m to \ref{Figure 4}p show line scans of the Ni (red curve) and Ga (green curve) elemental distributions as a function of position along the NT diameter. The signals were extracted in the area enclosed by the dotted line and averaged along the NT long axis. These curves allowed us to quantify the thickness of the $\ce{Al2O3}$/Ni shells on the nanotemplates (Methods). Thicknesses of the top and bottom edge of the cross section are highlighted by yellow and gray shaded regions in Figure \ref{Figure 4}m to \ref{Figure 4}p. Similar analysis where acquired at random spots of five NTs of each deposition, giving averaged thicknesses of 11.5 $\pm$ 2.4 nm, 14.0 $\pm$ 2.1 nm , 29.4 $\pm$ 4.1 nm and 12.2 $\pm$ 1.5 nm for depositions prepared with $c = 1, 5, 7$ and $10$, respectively. Only for $c = 7$ did we observe that the Ni shell had a thickness that was comparable with the one deposited on the planar Si (100) substrate. In Figure~S2 we analyzed the step coverage as the ratio between sidewall and top-surface film thickness for two samples. The values were $\geq 88~\% $.

The growth rates of shells were found to vary as a function of $c$. We explain the different growth rates and surface roughnesses by different growth mechanisms. The higher the number of nickel oxide steps $c$ in the ALD sequence, the more the surface is oxidized, to which the hydrogen plasma pulse is applied. The drop in growth rate for $c>1$ in Figure~\ref{Figure 5}a might indicate that we moved away from the PE-ALD regime, for which the growth rate is enhanced by reaction mechanisms involving plasma species, and entered a thermal ALD regime, which often exhibits a smaller growth rate \cite{Sioncke2009,VanHemmen2007}. Furthermore, the decomposition of the metalorganic precursor is expected to be more pronounced on nickel than on nickel oxide \cite{Ospina-Acevedo2019}. Consequently the Ni-rich surface realized by $c=1$ promotes a larger growth rate.
The discrepancies in growth rates for planar films and NT shells are attributed to the recombination of plasma species by collisions on the enlarged surface area provided by the nanotemplates. The recombination probability of hydrogen plasma species is higher on metallic nickel \cite{Wood,Kariniemi2012} than on an oxidized surface\cite{Grubbs2006,Tserepi1994}. For $c=7$ we observed the lowest roughness in planar thin films and the highest growth rate on the high aspect ratio nanotemplates. We assume that in this deposition process a scenario was reached where, at each ALD cycle, a monolayer of nickel oxide was formed before the reducing hydrogen plasma step was inserted. For $c=10$, the increase in growth rate for the planar film (Figure~\ref{Figure 1}b) suggested that a larger nickel oxide thickness was realized compared to samples with $c=5$ and $c=7$, meaning that the duration of the hydrogen plasma pulse was no longer optimized to reduce the deposited oxide layer. For the same reason, we speculate that the subsequent annealing treatment in hydrogen may have been less effective on the reduction-induced crystallization process of sample $c=10$ and thereby explain the less intense peaks observed in the X-ray diffractogram of Figure~\ref{Figure 2}e.

\subsection{Magnetic properties of planar Ni thin films}
\begin{figure}[h!]
	\includegraphics[width=0.5\textwidth]{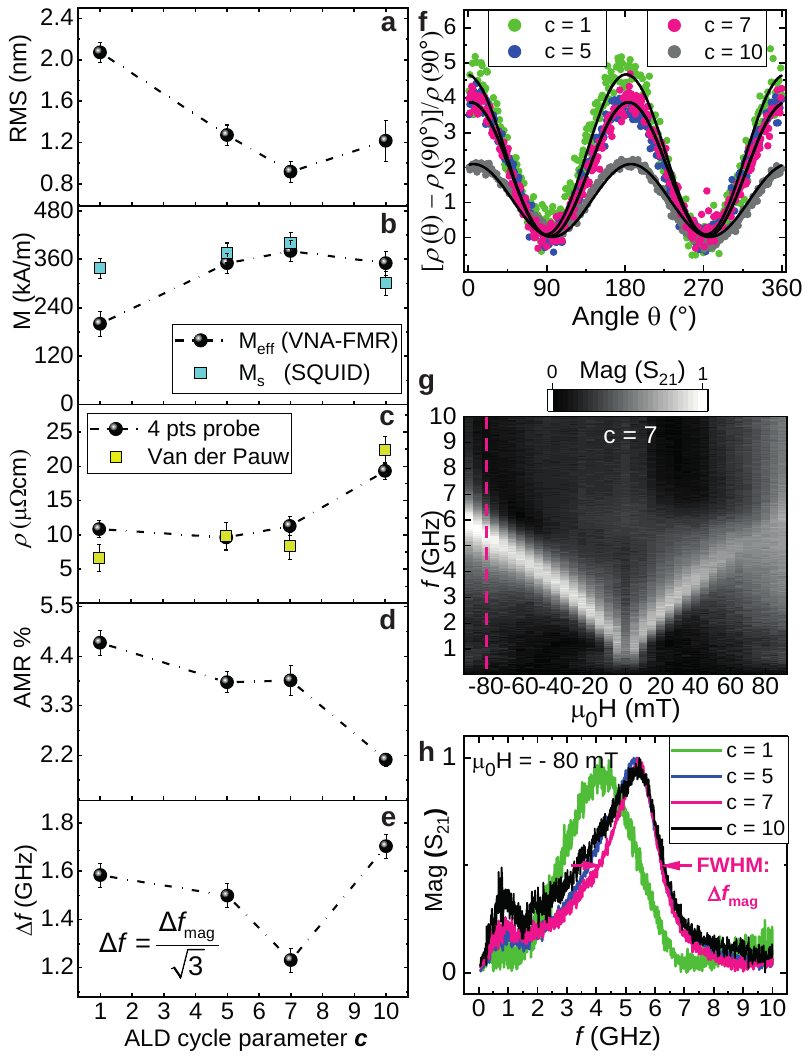}
	\caption{Physical properties of the Ni thin films as a function of $c$: (a) roughness, (b) effective and saturation magnetization $M_{\rm s}$ extracted from $M(H)$ curves (Figure S3b), (c) specific resistivity $\rho$ extracted by two independent experiments, (d) relative AMR effect, and (e) linewidth $\Delta f$. (f) Magnetoresistance of four thin-film samples (see labels) measured at room temperature for a rotating in-plane field of 80 mT. (g) Magnitude of the scattering parameter $S_{21}$ measured in an in-plane magnetic field by VNA-FMR on a thin film deposited with $c=7$. White indicates large absorption. (h) Spectra taken at -80~mT on Ni thin films prepared with different $c$ (see labels). For $c\geq 5$ the resonance frequencies are found to be the same for different $c$ but still the linewidth $\Delta f_{\rm mag}$ (indicated by horizontal arrows) varies as summarized in (e).}\label{Figure 5}
\end{figure}

In Figure~\ref{Figure 5}a to \ref{Figure 5}e (and Figure S3a and S3b) we summarize different physical properties of Ni thin films measured as a function of the growth parameter \textit{c}:
the rms roughness measured by atomic force microscopy (AFM) (Figure~\ref{Figure 5}a), the magnetization measured at room temperature by SQUID (Figure~\ref{Figure 5}b), the room temperature resistivity measured by either a four-probe configuration or a  van der Pauw configuration (Figure~\ref{Figure 5}c), the AMR effect (Figure~\ref{Figure 5}d) and the linewidth of the ferromagnetic resonance (FMR) (Figure~\ref{Figure 5}e). In Figure~\ref{Figure 5}f we depict the magnetoresistance measured on planar Ni thin films when an applied in-plane magnetic field $\mu_0H= 80$~mT was rotated (the angle $ \theta $ is defined between the current direction and $\mathbf{H}$). We display the specific resistivity $\rho(\theta)$ in terms of $ \dfrac{\Delta \rho(\theta)}{\rho} $ = $\dfrac{\rho(\theta) - \rho(90^{\circ})}{\rho(90 ^{\circ})}$. We observe a $cos^{2}(\theta)$ dependency as expected for the AMR effect. Depending on $c$ we find maximum values of the relative AMR effect $ \dfrac{\Delta \rho(\theta)}{\rho} $ ranging from 4.7 down to 2.1 (Figure~\ref{Figure 5}d). From field-dependent FMR measurement (Figure~\ref{Figure 5}e) we extracted the effective magnetization $M_{\rm eff}$ by fitting the Kittel formula \cite{Kittel1948} to the observed resonance frequencies (white branch in the color-coded spectra shown in Figure~\ref{Figure 5}g). The spectra were taken by inductive measurements using a vector network analyzer (VNA) and reflected the scattering parameter $S_{21}$ (Methods). Values of $M_{\rm eff}$ are summarized in Figure~\ref{Figure 5}b. In Figure~\ref{Figure 5}h  we compare individual spectra taken at the same field of -80~mT for thin films deposited with different $c$. We observe a resonance frequency of 5.5 GHz  and a narrow linewidth for $c=7$. Extracted linewidths are summarized in Figure~\ref{Figure 5}e.

The narrowest FMR linewidth of $\Delta f$ = 1.23 GHz is registered for a thin film prepared with $c=7$. The same film exhibits the smallest rms roughness of 0.9~nm. We assume that the small roughness reduces the inhomogeneous linewidth broadening expected from two-magnon scattering induced by inhomogeneities and surface irregularities\cite{Tserkovnyak2005,Arias1999}.
Values of $M_{\rm eff}$ are found to be in good agreement with $M_{\rm s}$ measured by SQUID (Figure S3b) in an in-plane magnetic field for $c\geq 5$. We measured a maximum effective magnetization $M_{\rm eff}$ of 380 kA/m for the sample deposited with $c=7$. Within the error bar it agrees with an $M_{\rm s}$ of 400 kA/m measured by SQUID on the same sample. Consistent values indicate that additional magnetic anisotropies (magnetocrystalline anisotropy, surface anisotropy) do not play a significant role. The experimentally observed values are about 20$~\% $ smaller than the reported saturation magnetization of 490 kA/m of single-crystalline nickel \cite{Crangle1971}. For $c=1$, $M_{\rm eff}$ is found to be smaller than M$_{\rm s}$ by about 130 kA/m. The discrepancy indicates an out-of-plane anisotropy present in this sample.
The lowest values of resistivity $\rho$ that we measured on thin films with $c=1$ and $c=7$ amounted to about 7 to 8.4 $\mu\Omega$cm. They were only slightly larger than the resistivity of 6.9 $\mu\Omega$cm reported for bulk Ni at room temperature \cite{Dean1999}. For $c=10$ we found $ \rho $ $\sim$ 20 $\mu\Omega$cm. In the same sample we find the smallest AMR value of 2.1~$\% $. We attribute the increased resistivity and the small AMR value for $c=10$ to inhomogeneities due to non-optimized ALD deposition.

\subsection{Transport properties of Ni nanotubes}

Nickel NTs of the different sets of depositions were extracted from the nanotemplate substrates and transferred to either Si or glass substrates for further nanolithography and the integration of metallic leads. Then we performed magneto-transport and spin-dynamic experiments on individual NTs. In the following we report on data obtained on samples S1 and S2 (Tab.~\ref{Tab1}).
\begin{table}
  \caption{Geometrical parameters of NTs prepared with $c=7$. S1 (S2) was used of magneto-transport (spin-dynamics) measurements.}
  \begin{tabular}{llll}
    \hline
    Sample  & $L$ ($ \mu $m) & $D_{out}$ (nm) & $t$ (nm) \\
    \hline
     S1 & 14.5 $ \pm $ 0.1 & 285 $ \pm $ 10 & 29.4 $ \pm $ 4.1\\
     S2 & 14.5 $ \pm $ 0.1 & 280 $ \pm $ 20 & 29.4 $ \pm $ 4.1\\
    \hline
  \end{tabular}\label{Tab1}
\end{table}
\begin{figure}[h!]
	\includegraphics[width=0.5\textwidth]{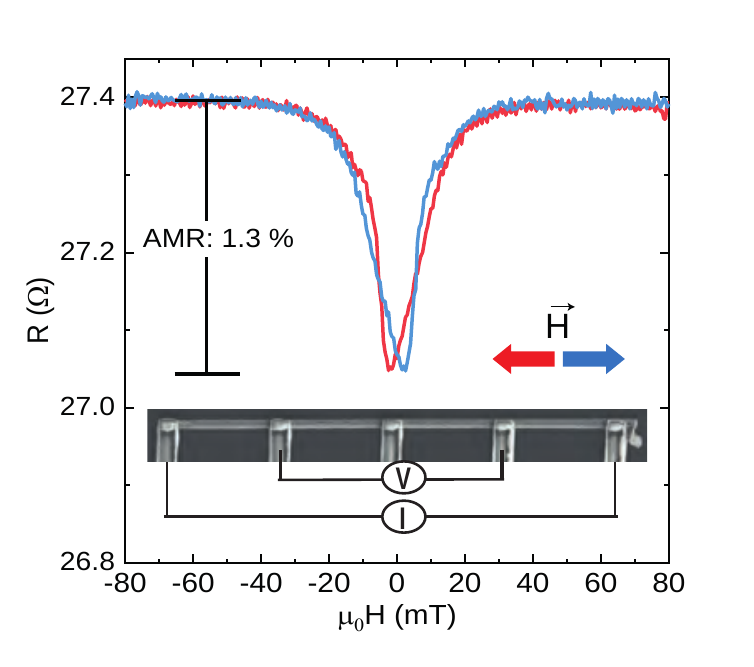}
	\caption{Magnetoresistance of nanotube S1 measured as a function of a static magnetic field applied along the long NT axis. The inset shows an SEM micrograph of S1 contacted by 5 metallic leads made from a Au thin film via lift-off processing. Voltage ($V$) and current ($I$) probes are indicated.}\label{Figure 6}
	
\end{figure}
In Figure \ref{Figure 6} and Figure S4, we report the field-dependent resistance $R$ measured on different nanotubes. The data of nanotube S1 deposited with $\textit{c} = 7$ is displayed in Figure~\ref{Figure 6}. The two voltage probes were separated by 7 $\mu$m. The field was applied long the long axis. While decreasing the field from 80 mT (red curve), the magnetoresistance remains constant at about 27.4~$\Omega$ until $\approx 30$ mT. From $R$ we calculate a specific resistivity of 7.8 $\pm$ 2.8 $\mu\Omega$cm. This value is only about 13 $\%$ larger compared to the resistivity of bulk nickel. For further decreased $H$, the resistance decreases to a minimum value of 27.05~$\Omega$ at -2 mT. The maximum resistance is restored again for $\mu_0H\leq -30$~mT and remains constant until $-80$ mT. The blue curve is acquired with increasing field and shows hysteretic behavior near zero field. Considering the long axis of the NT to be the easy axis, a field of 80 mT is sufficient to saturate the spins along the current direction. This configuration corresponds to the scenario of maximum electron scattering for the AMR effect and the resistance is at its maximum value. For fields below 30 mT, the resistance decreases, which we attribute to the onset of reversal of the NT's magnetization. The reduced resistance suggests that spins tilt away from the current direction. For nanotubes of similar geometrical parameters the reversal of magnetization is predicted to occur via vortex domain wall formation \cite{Landeros}. Correspondingly spins would be aligned in an azimuthal direction around the NT, thereby taking an angle of 90 deg with respect to the current flow direction. This configuration corresponds to the scenario of minimum electron scattering for the AMR effect and is consistent with the observed reduction of $R$ in the reversal regime. The relative variation of $R$ corresponds to an AMR effect of about 1.3~$\%$ at room temperature in small fields. This value is larger by about 30~$\%$ compared to the best low-field AMR effect of Ni NTs reported previously (see the supplementary information of Ref.~\cite{Ruffer2014}).

\subsection{Spin dynamics in Ni nanotubes}

\begin{figure}[h!]
	\includegraphics[width=1\textwidth]{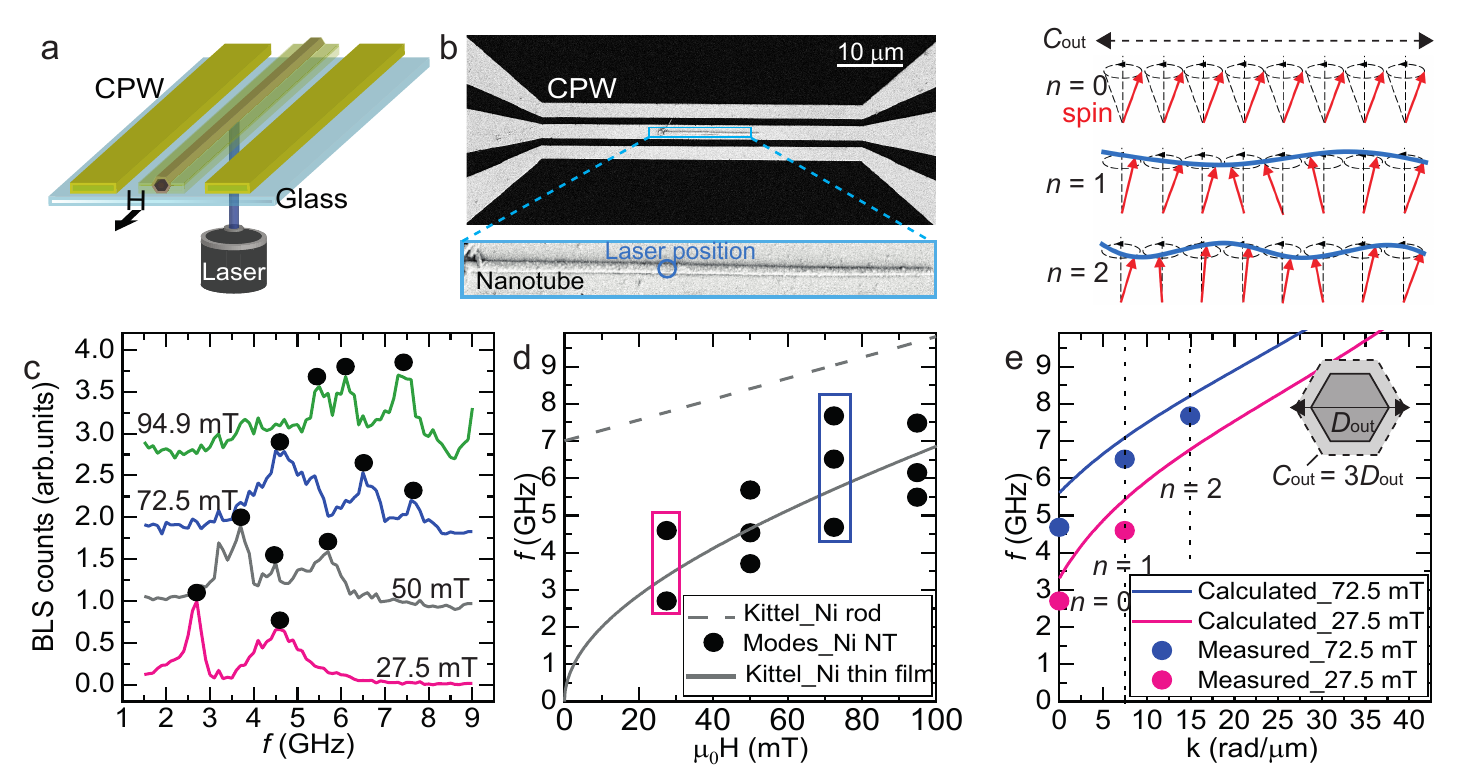}
	
	\caption{ (a)Sketch of the excitation-detection scheme to probe dynamic magnetization precession in individual NTs. The NT is covered by a coplanar waveguide (CPW) used for excitation. Laser light is focused through the transparent substrate onto the NT and the back reflected beam is analyzed by an interferometer. The field $H$ is collinear with the long axis of the NT. (b) SEM micrograph of the gold CPW laying on the investigated Ni nanotube S2. Inset displays a magnified image of S2 under the signal line. (c) BLS spectra detected on S2 at different applied static magnetic fields. Black circles indicate peaks attributed to resonance frequencies (eigenmodes). (d) Extracted eigenmodes plotted as a function of the field and compared with the resonance frequency given by the Kittel equation \cite{Gurevich96} for a nickel rod (dotted gray line) and a nickel thin film (continuous gray line) using $M_{\rm s} = 380$~kA/m and a gyromagnetic ratio of 28.02 GHz/T. (e) Spin wave dispersion relations calculated for a 30 nm Ni thin film subjected to 27.5 mT (bottom line) and 72.5 m T (top line) by using the Kalinikos and Slavin formalism \cite{Kalinikost1986}. The resonance frequencies measured on S2 (boxes in d) are displayed as filled circles matching the color code of the calculated lines. The wave vector $k=k_n$ was estimated assuming azimuthal confinement of DE modes with $n=0,1,2$ as illustrated above.}\label{Figure 7}
\end{figure}
The spin dynamics of individual NTs deposited with $c = 7$ were studied after depositing them on a separate glass substrate (Figure \ref{Figure 7}a and Figure S5). Coplanar waveguides (CPWs) made from 120 nm thick Au were fabricated by lift-off processing such that a NT was fully covered by the signal line of the CPW (Figure~\ref{Figure 7}b). When a microwave current was applied to the CPW, a dynamic magnetic field $h_{\rm rf}$ was generated, which exerted a torque on the spins of the NT. To detect the induced spin precession we used microfocused BLS ($\mu$-BLS) \cite{Demidov2008IEEE}. We focused a laser beam through the glass substrate onto the NT from the backside. The backreflected light was analyzed by means of a Fabry-Perot interferometer for energy shifts due to inelastic scattering between the photon and the spin excitation. The chosen geometry with the NT covered by Au and the laser focused through the substrate prevented the NT from degrading due to the ambient atmosphere when heated by the focused laser beam. BLS measurements were performed in the central position of the NT. Figure \ref{Figure 7}c shows spectra obtained by $\mu$-BLS while sweeping the signal generator frequency from 1.5 to 9 GHz in steps of 0.1~GHz. A static magnetic field was applied along the long axis and varied in steps from 27.5 (bottom curve) to 94.9 mT (top curve). At each increment we detect more than one peak in the spectrum, indicating multiple resonant magnon modes (black circles) whose eigenfrequencies vary systematically with $H$. In Figure \ref{Figure 7}d we summarize the resonance frequencies as a function of field. For comparison we show with a gray dotted and solid line the field-dependencies of FMR frequencies of a cylindrical nickel rod and a nickel film, respectively, which were calculated according to the Kittel formula \cite{Kittel1946} with a gyromagnetic ratio of 28.02 GHz/T and $M_{\rm eff} = 380$~kA/m, i.e. the value measured on the planar ALD-grown Ni film. Considering the shape-dependent demagnetization effects, we expect that the resonance frequency for uniform precession of a tube (FMR) resides below the FMR frequency of the rod. Indeed, measured eigenfrequencies of the NT fall below the dashed curve. The observed field-dependencies of measured eigenmodes are similar to the slopes of the calculated curves (broken and dashed lines). In the following, we speculate on the origin of the multiple observed eigenmodes. We assume that the peculiar tubular shape of the NT gives rise to standing spin waves with wavelengths $\lambda_n$ fulfilling the constructive interference condition $n\times \lambda_n = 3\times D$ along the azimuthal direction, where $n$ is an integer number ($n=0,1,2,...$) and $3\times D$ is the circumference $C$ of the hexagonal tube \cite{Ruffer2012}. The allowed wave vectors amount to $k_n = 2\pi/\lambda_n= n\times 2\pi/ C $. The mode with $n=0$ represents uniform spin precession. The wave vectors $\mathbf{k}_n$ with $n\geq 1$ point in azimuthal directions. They are orthogonal to the direction of magnetization $\mathbf{M}$ of the NT which was found to be aligned with $\mathbf{H}$ for $\mu_0H\geq 30$~mT (Figure~\ref{Figure 6}). In such a configuration the eigenfrequency of a spin wave is expected to increase with increasing $k$ (Damon-Eshbach mode configuration). In Figure~\ref{Figure 7}e we plot the peak frequencies observed at 27.5 mT and 72.5 mT in that for each field value we attribute wavevectors $k_n$ with increasing $n$ to modes with increasing resonance frequency. The solid lines shown in Figure~\ref{Figure 7}e reflect the spin wave dispersion relations \textit{f(k)} calculated for a 30 nm-thick Ni film according to the formalism provided by Kalinikos and Slavin \cite{Kalinikost1986}. The experimental values follow the slope of $f(k)$ expected for a thin film. However, the observed eigenfrequencies are smaller than the calculated ones at a given field. The remaining discrepancy could most likely be attributed to the different dynamic demagnetization effect of the tubular geometry compared to the planar film and the increased temperature of the NT in the laser focus. As the magnetization of Ni decreases with increasing temperature, lower resonance frequencies would indeed be expected. The results depicted in Figure~\ref{Figure 7} show that the Ni NTs form spin-wave nanocavities which impose discrete wave vectors and confine GHz microwave signals on the nanoscale. They can be produced in large ensembles of vertically aligned 3D nanomagnetic structures (Figure~\ref{Figure 3}).

\subsection{Conclusions}
Combining cycles of thermal ALD of nickel oxide with a single pulse of plasma-assisted hydrogen reduction, we both obtained the conformal growth typical of thermal ALD and exploited the low temperature reduction realized by a plasma treatment. The optimized ALD sequence resulted in conformal coating of vertically aligned nanotemplates with Ni where the growth rate was similar to the growth rate of the planar films. We achieved a specific resistivity of Ni of about 8~$\mu\Omega$cm on both planar substrates and arrays of nanotemplates. The low-field AMR effect observed in the hysteretic regime of a nanotube amounted to 1.3~$\%$ at room temperature. The spin-wave damping was low allowing us to detect several standing spin wave modes which fulfilled the constructive interference condition in azimuthal direction of a nanotube. The Ni thin films and nanotubes thereby exhibited physical properties which make them promising for functional spintronic elements and magnonic applications in 3D device architectures.

\section{Methods}
\subsection{Plasma enhanced atomic layer deposition}
Si (100) wafers were cleaved in pieces of circa 2 cm x 1cm and employed as substrates for the deposition of planar Ni thin films. The Si substrates were cleaned using isopropanol, DI water and dried by blowing nitrogen. GaAs NWs were grown in a moelcular-beam epitaxy reactor as previously reported \citep{Uccelli2011} and used as nanotemplates for nanotubes depositions. They were inserted in the ALD chamber without previous treatment. Nickel growth experiments were performed in a hot wall Beneq TFS 200 ALD reactor, operated at a pressure of 4-5 mbar, under a 100 sccm costant flow of ultrahigh purity nitrogen, used both as carrier and purge gas. Prior to Ni growth, the Si (100) wafer pieces and GaAs nanowires were coated with 5 nm of ALD \ce{Al2O3} in the same ALD chamber, by alternating trimethylaluminium (\ce{TMA}) and \ce{H2O} pulses for 65 cycles.
The growth of nickel was performed using nickelocene (\ce{NiC5H5}, \ce{NiCp2})as a precursor, water as the oxidant agent and an in-cycle plasma enhanced reduction step with hydrogen. Nickelocene and water were stored in stainless steel containers respectively at 80 $^\circ$ C and room temperature to exploit their vapor pressure.
The plasma was generated in a RF parallel plate system where one electrode with shower head was powered at 150 W, the other was grounded and hosted the sample substrates.
An ALD cycle consisting of $c$ steps of \ce{NiCp2}/\ce{H2O}
leading to the formation of nickel oxide, followed by one step of plasma hydrogen, reducing nickel oxide to metallic nickel, was repeated $n$ times. The ALD sequence can be summarized as follow: $[[\ce{NiCp2}/{\rm purge}/\ce{H2O}/{\rm purge}]\times c + \ce{H2}~{\rm plasma/purge} ]\times n$. The pulse duration of nickelocene and water in the \ce{NiCp2}/\ce{H2O} part of the sequence was set as 2 and 6s, respectively, and were followed by purge steps of 4 and 8 s. Hydrogen plasma was supplied for 4 s, followed by a nitrogen purge of 8 s. The thickness of the deposited thin films was measured imaging the films in cross section by SEM. The growth rate was calculated dividing the thickness by the total number of \ce{NiCp2} pulses $c\times n$ .
The first set of depositions was performed at different chamber temperatures of the range 160 - 175 $^\circ$ C, keeping \textit{c} = 1 and n = 1000.
A second set of samples was deposited at $T= 170~^\circ$ C to assess variations of growth rate on the number of \ce{NiCp2}/\ce{H2O} steps $c$, by keeping $c\times n=1000$.
To achieve thin films with thickness of around 30 nm, a third set of depositions was carried out where the ALD cycle number $n$ was set as 1000, 475, 350, 175 for \textit{c} set as 1, 5, 7, 10, respectively. After nickel deposition, the samples were annealed in the same ALD chamber at 350 $^\circ$ C, for 5 hours, under a mixture of hydrogen and nitrogen gas. No significant thickness variations were registered in the planar thin films as a consequence of the annealing treatment.

\subsection{Structural characterization and chemical analysis of the thin films and nanotubes}
We report the properties of the annealed thin films. Their morphologies and thicknesses were investigated by scanning electron microscopy (SEM) and atomic force microscopy (AFM). X-ray diffraction spectra were recorded in the glancing
incidence mode on a Malvern Panalytical(Empyrean model) diffractometer with incidence angle of 2$^\circ$.
The morphologies of the annealed nanotubes was investigated by both SEM and transmission electron microscopy (TEM), chemical element distribution was examined by scanning transmission electron microscopy (STEM) combined with energy dispersive X-ray spectroscopy (STEM-EDS). The TEM and STEM experiments were carried out using an FEI Talos electron microscope operated at 200 kV. The thicknesses of the NTs were extracted from EDS elemental 2D maps of Ga and Ni using Velox, which were analyzed along specific lines cutting the NT perpendicularly to the length.  Line scans were smoothened by applying a Gaussian blur (sigma~$=1.0$) on the maps. The width of the line was 15-20 nm. Moving from the inner GaAs core towards the external NT radius, we assumed the ALD \ce{Al2O3}/Ni shell to start at the point where the counts of the Ga element distribution (green curve) vanished to zero and to end at the point where the counts of the Ni element distribution (red curve)vanished to zero. We subtracted a nominal thickness of 5 nm ascribable to the \ce{Al2O3} layer to obtain the Ni thickness.

\subsection{Investigation of physical properties}
The static magnetic properties of the thin films were assessed using a SQUID system (MPMS-5, Quantum Design) operated at room temperature (300 K).
Broadband spectroscopy of the deposited thin films was performed using a a vector network analyzer (VNA)to sweep the frequency and record the ferromagnetic resonance (FMR) absorption spectra. The thin films were positioned on top of a CPW connected by microwave tips to a vector network analyzer (VNA). The 2-port VNA allowed us to generate a microwave magnetic field with frequencies ranging from 10 MHz to 10 GHz. The applied microwave current generated an in-plane rf-magnetic field perpendicular to the long axis of the CPW. The microwave with a power of $-15~$dBm was applied at the port 1 of the CPW in order to excite magnetization precession. The precession-induced voltage was detected at port 2 via reading the scattering parameter $S_{21}$ where the numbers 2 and 1 in the subscript denote the detection and excitation port. An external magnetic field $\mu_0H$ was swept from 90 mT to $-90~$mT along the CPW's long axis. The linewidth of the resonant trace acquired at each field was divided by $\sqrt{3}$ to obtain, with good approximation, the imaginary component of the FMR response.
Resistivity measurements were performed with a KLA Tencor OmniMap RS75 four-point resistivity meter. Anisotropic magneto resistance (AMR) measurements were carried out in Van der Pauw four-point configuration \cite{Kateb2019}. The samples were bonded into a chip-carrier adapted for a room temperature probe station, equipped with a custom built 2D vector magnet assembly that allowed us to vary the in-plane applied field $H$ under an angle $\theta$. AMR measurements were performed at room temperature applying a current of 2 $\mu$A and a static in-plane magnetic field of 80 mT applied at an angle varying from 0 to 360 deg.

\subsection{Experiments on nanotubes with integrated leads and waveguides}
The Ni nanotubes were transferred trough an isopropyl alcohol solution on a 4-inch Si(100) wafer covered with 200 nm thick \ce{SiO2}~for the fabrication of metallic leads and on a 4-inch fused silica wafer for the fabrication of integrated coplanar waveguides (CPWs). On both wafers pre-patterned gold alignment markers were fabricated by photolithography via a custom developed software previously reported \cite{Ruffer_T} to precisely localize randomly oriented nanotubes and generate the electron beam lithography pattern for integrated leads and CPWs. The metallic leads (CPWs) were prepared by electron beam lithography and a following sputtering (evaporation) of 5 nm Ti/ 400 (120) nm Au film. The CPWs' dimensions were chosen to enable impedance matching. The signal line, having a width of 2.0 $\pm$ 0.1 $\mu$m was separated by gaps of 1.1 $\pm$ 0.1$\mu$m width from the 2$\mu$m wide ground lines. Magneto-transport properties of nanotubes were investigated at room temperature employing a probe station. The resistances were measured by 4 probe measurements with a magnetic field applied in parallel to the tube axis. Resistivities were calculated as $\rho = \dfrac{R \times A}{l}$ , where $l$ is the separation between the voltage leads and $A$ is the cross-sectional area $A$ of the shell around a NW. Correspondingly, $A$ was calculated as the difference between an external hexagonal area with the hexagon's long diagonal $D_{\rm out}$ (measured by SEM) and an internal hexagonal area with diameter $D_{\rm in}$ (taken as $D_{\rm out}$ - $2\times t$, with $t$ being the thickness of the Ni shell).\\
Spin wave eigenmodes were detected via Brillouin light scattering (BLS) microscopy at room temperature \cite{Demidov2008IEEE}. A coplanar microwave waveguide (CPW) was fabricated on the glass substrate on top of the NT and oriented in a way that the signal line covered the NT along its long axis. The end of the CPW was electrically bonded to a printed circuit board, which was connected to a signal generator (Anritsu MG3692C) applying a microwave current. The corresponding magnetic microwave field excited spin precession in the NT at a fixed frequency. The frequency was swept from 1.5 to 9 GHz. A monochromatic laser with a wavelength of 473 nm and power of 0.8 mW was focused through the sample backside to a diffraction limited spot using a specially corrected $100\times$ objective lens with a large numerical aperture of NA = 0.85. The recorded BLS signal is proportional to the square of the amplitude of the dynamic magnetization at the position of the laser spot. The sample was mounted on a closed loop piezo stage which allowed a precise localization of the NT spots addressed in the experiments. The power was such that spin precession was excited in the linear regime. A magnetic field was applied parallel to the NT long axis via a permanent magnet.

\begin{acknowledgement}

Funding by the German Science Foundation DFG via GR1640/5-2 in SPP1538 “Spin caloric transport” and SNF via grants 163016, BSCGI0{\_}157705, and NCCR QSIT is gratefully acknowledged. We thank G{\"o}zde T{\"u}t{\"u}nc{\"u}oglu and Didier Bouvet for excellent experimental support.

\end{acknowledgement}

\begin{suppinfo}

Further information and data are contained in the supporting information. Data are available on zenodo.

\end{suppinfo}


\bibliography{library_corrected}



\section*{Supplementary material}

Supporting information

**Plasma-enhanced atomic layer deposition of nickel nanotubes with low resistivity and coherent magnetization dynamics for 3D spintronics**

M. C. Giordano[1], K. Baumgaertl[1], S. R. Escobar Steinvall[2], J.Gay[1], M. Vuichard[1], A. Fontcuberta i Morral[2,3], and D. Grundler[1,4]

[1] Ecole Polytechnique Federale de Lausanne, School of Engineering, Institute of Materials, Laboratory of Nanoscale Magnetic Materials and Magnonics, 1015 Lausanne, Switzerland

[2] Ecole Polytechnique Federale de Lausanne, School of Engineering, Institute of Materials, Laboratory of Semiconductor Materials, 1015 Lausanne, Switzerland

[3] Ecole Polytechnique Federale de Lausanne, School of Basic Sciences, Institute of Physics, 1015 Lausanne, Switzerland

[4] Ecole Polytechnique Federale de Lausanne, School of Engineering, Institute of Microengineering, 1015 Lausanne, Switzerland

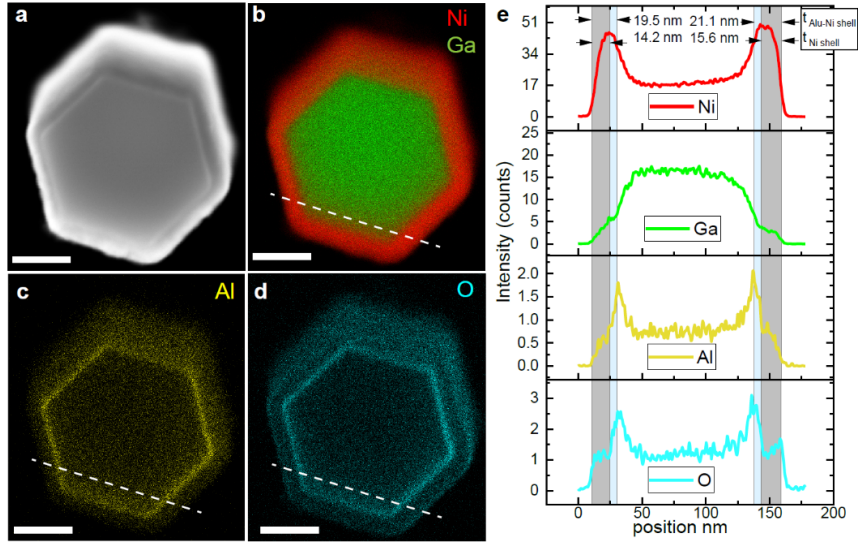

**Figure S1**. (a) HADDAF and elemental distribution of (b) Ni, Ga, (c) Al and (d) O of a cross section of a Ni shell deposited on a GaAs nanowire covered with a 5 nm thick $Al_2O_3$ layer (scale bars: 100 nm). (e) Elemental analysis along the white dotted lines. The asymmetric appearance of the shell thickness is due to the misalignment of the nanotube's long axis with the electron beam.

**Table S1**: Geometrical parameters employed to determine the aspect ratio (AR) of the nanowires templates. The minimum, maximum and average spacing among nanowires are indicated as $w_{min}$, $w_{max}$ and $w_{ave}$, respectively. $L_{max}$ represents the maximum length of the nanowires.

| Deposition set | $w_{min}$ (μm) | $w_{max}$ (μm) | $w_{ave}$ (μm) | $L_{max}$ (μm) | AR = $L_{max}$ / $w_{ave}$ |
|---|---|---|---|---|---|
| **c = 1** | 0.43 ± 0.1 | 0.78 ± 0.1 | 0.61 ± 0.25 | 15.0 ± 0.1 | ~ 25 : 1 |
| **c = 5** | 0.17 ± 0.1 | 1.00 ± 0.1 | 0.54 ± 0.59 | 8.0 ± 0.1 | ~ 15 : 1 |
| **c = 7** | 0.88 ± 0.1 | 0.06 ± 0.01 | 0.49 ± 0.58 | 15.0 ± 0.1 | ~ 31 : 1 |
| **c = 10** | 0.50 ± 0.1 | 0.20 ± 0.1 | 0.35 ± 0.21 | 7.0 ± 0.1 | ~ 20 : 1 |

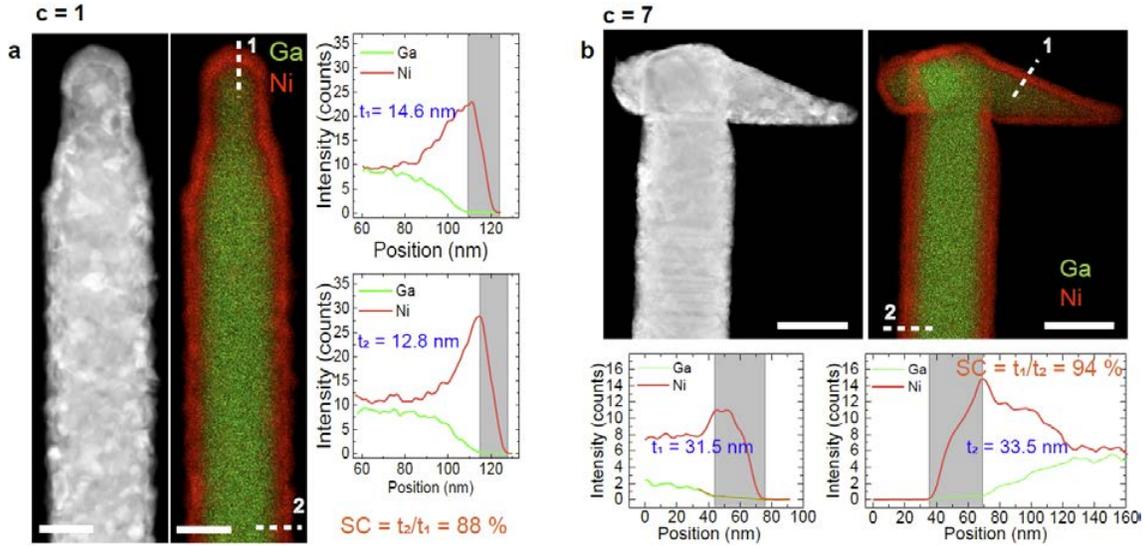

**Figure S2.** STEM-EDS images taken on ALD-grown Ni shells of depositions sets with (a) $c = 1$ (scale bar: 100 nm) and (b) $c = 7$ (scale bar: 200 nm) deposited on GaAs nanowires after depositing 5 nm of $Al_2O_3$. The core/shell systems were annealed at 350 °C. The ratio $t_2/t_1$ of thicknesses $t_1$ and $t_2$ taken at positions 1 and 2 respectively, provide the step coverage ratio SC. The two positions are separated by about 900 nm. SC is given in the panels.

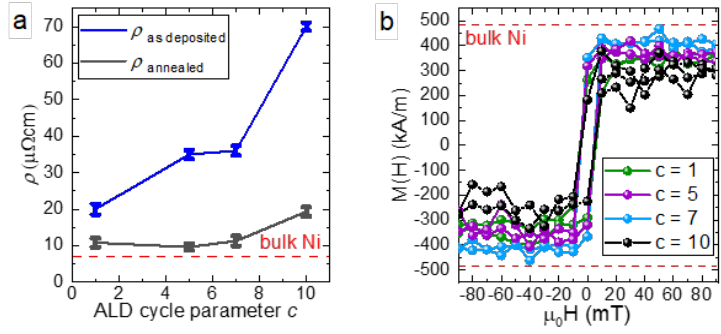

**Figure S3**. (a) Values of resistivity of the four planar ALD-grown Ni films as deposited on a Si substrate, displayed as a function of the ALD cycle parameter $c$ (blue curve). The values are compared with the ones reported in the main manuscript for the corresponding annealed samples (dark gray curve). The resistivity was measured at room temperature in a 4-point probe configuration. (b) Magnetic hysteresis curves $M(H)$ of the annealed planar ALD Ni films. Ni films grown with $c = 5$ and $c = 7$ exhibit the largest saturation magnetization $M_s$ extracted at 85 mT. Values are reported in Fig. 5 of the main manuscript. The data were obtained at room temperature using a SQUID magnetometer with the external magnetic field $H$ applied in the plane of the films.

**Table S2:** Parameters of ALD-grown Ni NTs for which measured data are reported in Figures S4 and S5.

| Sample | L ($\mu$m) | $D_{out}$ (nm) | t (nm) | Deposition set |
|--------|------------|----------------|--------|----------------|
| s_1 | 14 ± 0.1 | 260 ± 10 | 11.5 ± 2.4 | $c = 1$ |
| s_2 | 14.4 ± 0.1 | 220 ± 10 | 13.7 ± 1.0 | $c = 5$ |
| s_3 | 5.0 ± 0.1 | 265 ± 10 | 29.4 ± 4.1 | $c = 7$ |
| s_4 | 13.7 ± 0.1 | 260 ± 10 | 29.4 ± 4.1 | $c = 7$ |

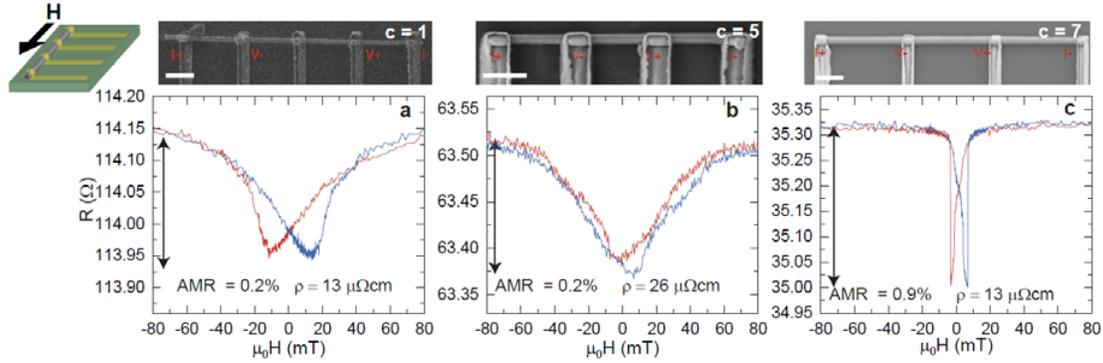

**Figure S4.** Magnetoresistance $R(H)$ measured on the Ni NTs (a) s_1, (b) s_2 and (c) s_3 described in Table S1 and corresponding SEM images. The resistance was measured at room temperature with the external magnetic field $H$ applied along the long axes (sketched on the left), by using the metallic leads sketched in the SEM images. Relative resistance changes (AMR) between maximum and minimum values of $R$ are stated in each figure. From values $R$ measured at 80 mT we calculated the specific resistivities $\rho$ as stated in the figures. Scale bars are 1 m. The core/shell systems were annealed at 350 °C.

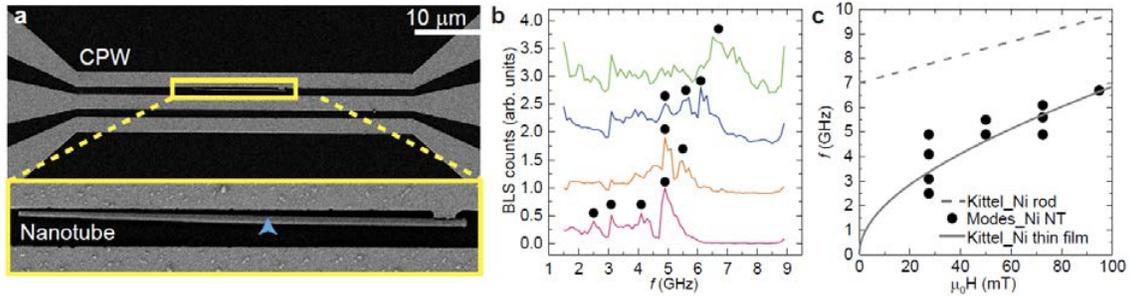

**Figure S5.** (a) SEM image of an individual nanotube (s_4) positioned in the gap of a coplanar waveguide (CPW) between signal and ground lines. The nanotube was grown with $c = 7$. (b) BLS data obtained when focusing the laser onto the center of the nanotube and exciting spin-precessional motion by microwave magnetic fields with frequencies between 1.5 and 9 GHz. We show spectra taken at fields 27.5, 50, 72.5 and 94.9 mT (from bottom to top) applied along the long axis. We mark peaks that we identify as magnon resonances with filled circles. (c) Summary of resonance frequencies extracted from (b) plotted as a function of applied field $H$. We compare the resonance frequencies with the field-dependent resonance frequencies expected for a planar Ni film (solid line) and a Ni nanorod with a diameter consistent with the nanotube (broken line). The resonance frequencies are grouped close to the solid curve consistent with the nanotube discussed in the main text.



\end{document}